\documentclass[11pt]{article}
\usepackage{mathrsfs}
\usepackage{amsmath}

\global\arraycolsep=1pt \oddsidemargin.20in \evensidemargin.5in
\usepackage[Symbol]{upgreek}

\usepackage{bbm}
\usepackage{dsfont}
\usepackage{amssymb}
\usepackage{textcomp}

\newcommand{\CR}{\nonumber \\*}

\DeclareMathAlphabet{\mathpzc}{OT1}{pzc}{m}{it}

\usepackage[colorlinks=true,backref=true,
linkcolor=black,anchorcolor=black,citecolor=black,
filecolor=black,menucolor=black,pagecolor=black,urlcolor=black]{hyperref}

\def\z{\zeta}

\def\be{\begin{equation}}
\def\ee{\end{equation}}
\def\bea{\begin{eqnarray}}
\def\eea{\end{eqnarray}}
\def\bdis{\begin{displaymath}}
\def\edis{\end{displaymath}}
\def\corr{$\clubsuit$}
\def\nn{\nonumber}

\begin{document}
\allowdisplaybreaks[1]
\renewcommand{\thefootnote}{\fnsymbol{footnote}}
\def\corr{$\spadesuit $}
\def\trefle{ $\clubsuit$}
\renewcommand{\thefootnote}{\arabic{footnote}}
\setcounter{footnote}{0}
 \def\stop{$\blacksquare$}
\begin{titlepage}
\null
\begin{flushright}
CERN-PH-TH/2010 \\
{CBPF-NF-005/11}
\end{flushright}
\begin{center}
{{\Large \bf
Twisted Superalgebras and Cohomologies of the\\
$N=2$ Superconformal Quantum Mechanics
}}
\lineskip.75em \vskip 3em \normalsize {\large Laurent
Baulieu$^{ \dagger\ddagger}$\footnote{email address:
baulieu@lpthe.jussieu.fr}
and
 Francesco Toppan$^{*}$\footnote{email address:
 toppan@cbpf.br}
\\
\vskip 1em
  $^{\dagger}${\it Theoretical Division CERN }\footnote{ CH-1211
  Gen\`eve, 23,
  Switzerland }
\\
$^{\ddagger}${\it LPTHE
Universit\'e Pierre et Marie Curie }\footnote{ 4
place Jussieu,
F-75252 Paris
Cedex 05, France}
\\
$^{* }${\it CBPF, Rio de Janeiro }\footnote{ {Rua Dr. Xavier
Sigaud 150, cep 22290-180, Rio de Janeiro, RJ, Brazil}}
 }

\vskip 1 em
\end{center}
\vskip 1 em
\begin{abstract}
We prove that the invariance of the $N=2$ superconformal quantum mechanics is controlled by subalgebras of a given twisted superalgebra made of $6$ fermionic (nilpotent) generators and $6$ bosonic generators (including a central charge). The superconformal quantum mechanics actions are invariant under subalgebras of this quite large  twisted superalgebra. They are in fact fully determined by a subalgebra with only $2$ fermionic and $2$ bosonic (the central charge and the ghost number) generators.  The invariant actions are $Q_i$-exact,  with a $Q_{i'}$-exact ($i'\neq i)$ antecedent for some of its $6$ fermionic generators. It follows that  the superconformal quantum mechanics  actions with  Calogero potentials   are  uniquely determined even if, in its bosonic sector, the twisted superalgebra does not contain the one-dimensional conformal algebra $sl(2)$, but only its Borel subalgebra.
The  general coordinate covariance of  the non-linear sigma-model for the   $N=2$ supersymmetric quantum mechanics  in a curved target space is fully implied only by its worldline  invariance  under a pair of the $6$  twisted supersymmetries.  
The transformation connecting the ordinary and twisted formulations of the $N=2$ superconformal quantum mechanics is explicitly presented.
 \end{abstract}

\end{titlepage}
\def\z{\bar z}
\def\bm{{\bar m}}
\def\bn{{\bar n}}
\def\bp{{\bar p}}
\def\bq{{\bar q}}
 \def\demi{\frac{1}{2}}
\section{Introduction.}

Both the ordinary and the conformal $N=2$ supersymmetric quantum mechanical models describe interesting dynamics. The $N=2$ supersymmetry has    applications to solvable potentials and the motion of particles in certain gravitational backgrounds \cite{fr}--\cite{gl}. It is  an  appropriate framework to describe topological invariants of various  target-spaces for the propagating bosons~\cite{wit}. The worldline
$N=2$ global supersymmetry has a natural extension as a superconformal algebra ($N=2$ SCA)
\cite{ap}.
~\par 
 The interpretation of the $N=2$ supersymmetric quantum mechanics as  a  topological quantum field theory derives from the invariance under a twisted superalgebra, whose nilpotent fermionic generators define a  BRST-type cohomology \cite{{bg},{brt},{bs},{bt}}. \par
We construct here a  twisted superalgebra, called ``twisted $N=2$ SCA",  acting on a set of  $(1,2,1)$ supermultiplets $\left(X_\mu(t); \Psi_\mu(t),\bar \Psi_\mu(t); b_\mu(t)\right)$. For any given $\mu$, $1\leq \mu\leq d$,  $X_\mu(t)$ is a
propagating boson,  $\Psi_\mu(t)$,
 $\bar \Psi_\mu(t)$  are its  two anticommuting  supersymmetric partners and    
$b_\mu(t)$  is a commuting auxiliary field. Such multiplets are called balanced because they contain  an equal number of bosonic and fermionic component fields.
For us the  expression ``twisted superalgebra"  just means that its fermionic generators are
all nilpotent; the relation  between the twisted $N=2$ SCA  and  the ordinary
$N=2$ SCA  is that 
 the invariance of an action under the 
twisted $N=2$ SCA implies the  invariance under the ordinary $N=2$ SCA and vice-versa. 
In fact, the construction extends that of \cite{bt}  from global supersymmetries to the case of supersymmetries whose generators carry an explicit dependence
on the $t$ coordinate.\par 
The ordinary one-dimensional $N=2$ superconformal algebra of the $(1,2,1)$ supermultiplet is the simple Lie superalgebra $sl(2|1)$ \cite{fss}, with $4$ bosonic and $4$ fermionic generators. 
It contains as a subalgebra, in its bosonic sector, the $sl(2)$ algebra which defines the one-dimensional Conformal Quantum Mechanics \cite{dff}.  \par
The existence    of the twisted $N=2$ SCA is interesting for at least two reasons. It   selects  invariants as local cocycles  and, furthermore, it contains quite  a small subalgebra which is sufficient to determine the full superconformal invariance. \par
To build   the twisted $N=2$ SCA, we can    construct at first
a  global twisted $N=2$ superalgebra with $4$ fermionic generators, acting on the $(1,2,1)$
supermultiplet,  
the two worldline ``twisted    scalar supersymmetries"  $Q$ and $\bar Q$ and the two   
``twisted  vector supersymmetries"  $Q_V$ and $\bar Q_V$. Then, one can complement these $4$  fermionic nilpotent generators by two extra nilpotent ones which carry the explicit
$t$-dependence, the ``twisted conformal supersymmetries" $Q_C$ and $\bar Q_C$. 

The ${\bf Z}_2$-graded   anticommutators of $6$ fermionic generators  close on a set  of $6$ bosonic
generators, one of them being    a central charge.      Due to the presence of the central charge, the 
twisted $N=2$ SCA is not a simple Lie superalgebra as the ordinary $N=2$ SCA. 
\par
The  generators of the 
twisted $N=2$ SCA can be   realized as $4\times 4$ supermatrices with $2\times 2$ even/odd blocks, whose entries are $t$-dependent differential operators. This provides a so-called $D$-module representation.\par
The twisted $N=2$ SCA induces the same invariant actions as the ordinary $N=2$ SCA. However, it does not contains  the conventional  conformal $sl(2)$ symmetry  as a bosonic
subalgebra. Rather,  only a Borel subalgebra of the conformal $sl(2)$ belongs to the twisted $N=2$~SCA. 
\par
We   found that imposing the invariance under  a small subalgebra is sufficient to
guarantee the invariance under the full twisted $N=2$ SCA and,  consequently, the ordinary $N=2$~SCA. 
\par
The minimal superalgebra which enforces the $N=2$ superconformal invariance is made of only $2$ fermionic generators (either  $Q$ and $\bar Q_C$ or  $\bar Q$ and $  Q_C$) and $2$ bosonic generators (the central charge $c$  with a fixed coefficient and the ghost number $N_{gh}$). Therefore, the whole superconformal invariance is obtained ``for free", with the extra generators regarded as ``accidental" symmetries of this minimal set-up.\par
This intriguing result offers a new perspective for investigating the conformal properties of the supersymmetric models, since it suggests that one can replace the demand of the conformal symmetry by that of a much smaller symmetry
\footnote{
These considerations have analogies in the $N=4$, $d=4$ super-Yang--Mills theory, where the superconformal Yang--Mills supersymmetry  with its $32$ supersymmetric generators is implied by a much smaller superalgebra, with only $4$ scalar twisted  nilpotent supercharges~\cite{bau}.}.\par
Conformally invariant topological  theories such as  the topological quantum field theory  toy model \cite{br}  and  the Calogero models  are   thus  very economically defined  by simply demanding the invariance under the  $2$ supercharges  $Q$ and $\bar Q_C$, instead of imposing   the complete  world-line superconformal symmetry.  New Lagrangians with higher-order interactions among Fermi fields, involving fields that are one-dimensional analogous of the Ramond fields, can also be  constructed.
\par
Further results in this paper can be summarized as follows.
The invariant actions $I$  are $Q_i$-exact,  with a local $Q_{i'}$-exact ($i'\neq i)$ antecedent for some of the $6$ fermionic generators of the twisted $N=2$ SCA ($I=\int dt  Q_{i} Q_{i'} Z_{i i'} $). \par
If one relaxes the condition of conformal invariance, the  general coordinate covariance of  the non-linear sigma-model for the  $N=2$ supersymmetric quantum mechanics  in a curved target space is fully implied by its worldline  invariance  under the action of only two of the above mentioned 6 supercharges, either  $Q$ and $\bar Q_V$ or 
  $\bar Q$ and $  Q_V$. (The invariance under the $2$ extra supercharges is automatically obtained). One can search for target metrics such that the action is   superconformally invariant \cite{pap}. \par
An invertible complex transformation relates the component fields and the free actions of the ordinary and the twisted $N=2$ superconformal algebra.\par
The paper is  organized as follows. In Section {\bf 2} we introduce the twisted $N=2$ superconformal algebra and discuss the relevant subalgebras. In section {\bf 3} we
investigate the different cohomologies which induce superconformally invariant actions (both free and in the presence of an interacting potential) in the case of a flat target-space. In Section {\bf 4} we discuss the implications of the cohomologies for the covariance of a curved target manifold. In Section {\bf 5} the construction of higher-order  Fermi interactions is pointed out. The explicit transformation relating the ordinary and the twisted formulations of the $N=2$ SCA is presented in Section {\bf 6}.
In the Conclusion we make a comparison between our observations concerning these simple supersymmetric quantum  mechanical systems and the intriguing results that have been recently observed in  higher dimensional supersymmetric quantum field theories.  We also discuss the future perspectives of our work.

\section{The twisted $N=2$ superconformal algebra.}
\def\intdt{\int  dt  }
\def\intdt{\int  dt  }

We recall at first some known facts.
 Let ${\cal M}_d$  be a $d$-dimensional manifold, locally parametrized by the $1\leq\mu\leq d$ coordinates $X^\mu$, with metric    $g_{\mu\nu}(X)$,
    Christoffel symbols $\Gamma_{\mu\nu,\rho}$ and   Riemann tensor 
    $R_{\mu\nu,\rho\sigma}$.
  The target-space reparametrization covariant action with worldline  $N=2$  supersymmetry is
expressed by the Lagrangian 
  \bea\label{cur1}
   {\cal L} 
   &= &      
     -\frac {1}{4}g_{\mu\nu}  
   \dot X^\nu\dot X^\nu
   +\bar \Psi^\mu(  g_{\mu\nu}  \dot\Psi^\nu +\Gamma _{\mu,\rho\sigma}\dot X^\rho\Psi^\nu)
   +\frac {1}{4}  R_{\mu\nu\rho\sigma }
   \bar \Psi ^\rho
   \Psi^\sigma
   \bar \Psi ^\nu
   \Psi^\nu.
   \eea
   $t$ parametrizes the worldline and $\dot \Phi \equiv\frac{d\Phi}{dt}$. The possibility of choosing any given  parametrization is obvious since we are working with a one-dimensional parametrization.\par
 The $t$-dependent coordinates $X^\mu(t) $  are bosons, while $\Psi^\mu(t)$ and $\bar\Psi^\mu(t)$ are fermions. Using an auxiliary field  $b^\mu (t)$, one can express ${\cal L}$ as  \cite{bs}
    \bea   \label{cur2}
    {\cal L}&=& g_{\mu\nu}  b^\nu
b^\mu
 +b^\mu( -g_{\mu\nu} \dot X^\nu  +   \Gamma _{[\mu,\rho]\sigma} \bar \Psi ^\rho
   \Psi^\sigma) +\partial_\rho g_{\mu\nu}\bar \Psi^\mu \Psi ^\rho X^\nu
   +\bar \Psi^\mu(  g_{\mu\nu}  \dot\Psi^\nu +\Gamma _{\mu,\rho\sigma}\dot X^\rho\Psi^\nu) .
         \eea
The general covariance  in the curved target-space with coordinates $X^\mu$  is explicit for the action~(\ref{cur1}). However, such an important invariance  is only enforced after the elimination from  the action~(\ref{cur2}) of the auxiliary fields $b^\mu$ via their algebraic equations of motion  $b_\mu=  g_{\mu\nu} \dot X^\nu  -  \Gamma _{[\mu,\rho]\sigma} \bar \Psi ^\rho
   \Psi^\sigma$. These equations show that   the ``on-shell" $b^\mu$-replaced fields are not vectors.
The action~(\ref{cur2}) is not invariant under target-space general coordinates transformations
due to the fact that it   inherently involves the  Christoffel symbols $\Gamma _{[\mu,\rho]\sigma}$ and that it is not possible to redefine the  $b^\mu$ fields in order to absorb this  dependence. On the other hand, when  the auxiliary  fields are present, all (twisted or untwisted) supersymmetry transformations close ``off-shell". After the elimination of the $b^\mu $ fields  via their equations of motion, the supersymmetries only close modulo  some fermionic equations of motion.  All this boils down to the fact that $X,\Psi,\bar\Psi,b$ is a balanced multiplet (for simplicity, from now on we drop, when not necessary, the $\mu$ suffix), while  $X,\Psi,\bar\Psi $ is an unbalanced multiplet,  that is it contains an unequal number of bosonic and fermionic component fields.
     These intriguing facts about what is happening  in the presence or after eliminating the auxiliary fields  are  however not troublesome when the fermionic twisted generators are realized as nilpotent generators  \cite{bs}. From the point of view of studying the world-line supersymmetry, the action~(\ref{cur2}) is more suitable.\par
 The balanced  quantum mechanical supersymmetric  multiplet  is thus made of $d$ independent $(1,2,1)$ supermultiplets, whose fields are target-space vectors
 \bea\label{341242}
  X^\mu(t) ,    \Psi  ^\mu(t) ,   { \bar \Psi  ^\mu(t) }  ,
b ^\mu(t) ,\ \ \  1\leq\mu\leq d.
\eea
 In the flat case the metric is   $g_{\mu\nu}(X) =\eta_{\mu\nu} $ and the Lagrangian is simply
given by
\bea {\cal L} &=& b^\mu
   \eta_{\mu\nu}  b^\nu
-b^\mu
   \eta_{\mu\nu}  \dot X^\nu
   +\bar \Psi^\mu  \eta_{\mu\nu}   \dot \Psi^\nu\sim
   -\frac{1}{4}
   \eta_{\mu\nu} \dot X^\mu \dot X^\nu
   +\bar \Psi^\mu  \eta_{\mu\nu}   \dot \Psi^\nu.
   \eea
 Two important  bosonic charges are conserved and compatible  with all fermionic transformations, the field dimension (also known as ``engineering dimension") and the ghost number.  Their values for the components of the balanced multiplet are, respectively, $(-\frac{1}{2}, 0,0,\frac{1}{2})$ and $(0, 1,-1,0)$. The action has  ghost number  zero and is dimensionless
if we assume the world-line parameter $t$ to possess the engineering dimension $-1$. \par
  In the following we will show the existence of other fermionic invariances of the free Lagrangian. We will check which interactions can preserve at least some of them.  

\subsection {Construction and presentation of the twisted $N=2$ SCA.}
 
 The $6$ nilpotent fermionic generators can be divided into one  pair of worldline  scalar twisted supersymmetry operators $Q,\bar Q$,
 one pair of worldline  vector  twisted supersymmetry operators $Q_V ,\bar Q_V$
 and one  pair of worldline  scalar  special twisted supersymmetry operators $Q_C,\bar Q_C$. 
$Q,\bar Q, Q_V ,\bar Q_V$ are constructed with the prescription of \cite{bt}, while 
$Q_C, \bar Q_C$ are determined by demanding explicit $t$-dependence and compatibility with
ghost number and engineering dimension.
 
These $6$ operators act on the component fields according to the transformations
      \bea\label{superalgebra}
\begin{array}{lll}
 QX  =\Psi    , &\quad & \bar QX  =\bar\Psi,\\
Q\Psi   =0    , &\quad & \bar Q\Psi   = -b,\\
Q\bar \Psi    = b    , &\quad & \bar Q \bar\Psi=0,\\
Qb  =0    , &\quad & \bar Qb  = 0.\\
&&\\
 Q_VX  = \bar \Psi    , &\quad & \bar Q_VX  =  \Psi,\\
Q_V\Psi   =-b+\dot X    , &\quad & \bar Q_V\Psi   = 0,\\
Q_V\bar \Psi    = 0    , &\quad & \bar Q_V \bar\Psi=b-\dot X,\\
Q_Vb  =\dot {\bar\Psi}    , &\quad & \bar Q_Vb  = \dot {\Psi}.\\
&&\\
 Q_CX  = t \Psi    , &\quad & \bar Q_CX  =  t \bar\Psi,\\
Q_C\Psi   =0   , &\quad & \bar Q_C\Psi   = -tb    +\bar\lambda X,\\
Q_C\bar \Psi    =tb    -\lambda X    , &\quad & \bar Q_C \bar\Psi=0,\\
Q_Cb  =\lambda  \Psi    , &\quad & \bar Q_Cb  = \bar\lambda  {\bar \Psi}.
\end{array}
\eea
Until now the real parameters $\lambda, \bar \lambda$ are arbitrary. We are however forced to
set $\bar\lambda=\lambda$ in order to eliminate unwanted anticommutation relations
(the presence of a $t$-multiplication operator)  arising from $\{Q_C , \bar Q_C \}$. This setting guarantees that $\{Q_C, \bar Q_C\}=0$.\par
  Therefore,   
the only non-vanishing anticommutators are
\bea\label{anticomm}
\begin{array}{ccc}
\{Q , Q_V\} =H,&\quad  &\{\bar Q ,\bar   Q_V\}= H,\\
\{Q,\bar Q_C\}=c,&\quad&\{\bar Q, Q_C\}=-c,\\
\{Q_C,Q_V\}=S,&\quad&\{\bar Q_C,\bar Q_V\}=\bar S,\\
\{ Q_V,\bar Q_C\}=Z,&\quad&\{ \bar Q_V, Q_C\}=\bar Z.
\end{array}
\eea
A  central charge $c$,
\bea
c&=&\lambda{\bf 1},
\eea 
has arisen from the anticommutators of $Q$ with $\bar Q_C$ and 
 $\bar Q$ with $  Q_C$.\par
  The action on the component fields of the bosonic operators $H, S, \bar S, Z, \bar Z$ is as follows 
 \bea\label{hss}
 H&=&\frac{d}{dt},\nonumber\\
S&=&   t \frac{d}{dt}   + \Delta, \nonumber\\
\bar S&=& -t\frac{d}{dt}  + \bar  \Delta,
\eea
and
\bea
&
\begin{array}{lll}
 ZX  =Z\bar\Psi=Zb=0    , & \quad& Z\Psi  = \bar\Psi,\\
\bar Z X=\bar Z\Psi=\bar Z b=0, &\quad& \bar Z \bar\Psi   = -\Psi,
\end{array}
&
 \eea
 where $\Delta$, $ \bar \Delta$ in (\ref{hss}) act as diagonal operators: 
      \bea&
\begin{array}{ll}
 \Delta X  =-\lambda X, \quad\quad&  \bar \Delta X    =  \lambda X,  \\
 \Delta b   =(1-\lambda)b,\quad\quad&  \bar \Delta  b= (\lambda-1) b,  \\
    \Delta \Psi= (1-\lambda)\Psi,\quad\quad&\bar \Delta \Psi =   \lambda \Psi, \\
 \Delta \bar \Psi=-\lambda\bar\Psi,\quad\quad&
 \bar \Delta\bar \Psi=(\lambda-1)\bar\Psi.
\end{array}&
\eea

The $12$ operators entering (\ref{anticomm}) are closed under (anti)commutation relations, so that \bea
 {\cal G}^\sharp\equiv\{Q,\bar Q, Q_V, \bar Q_V,Q_C,\bar Q_C, H, c, S,\bar S, Z,\bar Z\}\eea is a
Lie superalgebra. \par
The non-vanishing commutators involving the even operators of ${\cal G}^\sharp$ are 
\bea\label{comm1}
\relax [H,S]&=&[\bar S,H]=H,\nonumber\\
\relax[S,Z]&=&[\bar S, Z]=Z,\nonumber\\
\relax[\bar Z, S]&=&[\bar Z,\bar S]=\bar Z,\nonumber\\
\relax [\bar Z,Z]&=&S+\bar S.
\eea
The non-vanishing commutators between even and odd generators of ${\cal G}^\sharp$ are
\bea
&
\begin{array}{lll}\label{comm2}
\relax [H,Q_C]=Q,\quad &[H,\bar Q_C]=\bar Q,\quad& \\
\relax [S,Q]=-Q,\quad &[S,\bar Q_V]=-\bar Q_V,\quad& [S,\bar Q_C]=\bar Q_C,\quad\\
\relax [\bar S,\bar Q]=\bar Q,\quad &[\bar S,Q_V]=Q_V,\quad& [\bar S, Q_C]=-Q_C,\quad\\
\relax [Z,Q]=-\bar Q,\quad &[Z,\bar Q_V]=-Q_V,\quad& [Z,Q_C]=-\bar Q_C,\quad\\
\relax [\bar Z,\bar Q]=Q,\quad &[\bar Z, Q_V]=\bar Q_V,\quad& [\bar Z,\bar Q_C]=Q_C.\quad\\
\end{array}
&
\eea 

From the action of the ${\cal G}^\sharp$ operators on the component fields one can immediately write
down a $D$-module representation of ${\cal G}^\sharp$ in terms of $4\times4$ supermatrices, as displayed in the Appendix A.
\par
The  Lie superalgebra ${\cal G}^\sharp$  is compatible with the following assignment for the  scaling dimensions 
of  the component fields and of the generators (we set for the worldline coordinate $t$ the dimension $[t]=-1$). For the component fields we have
\bea
&[\Psi]=x+z,\quad [\bar \Psi]=x+1-z,\quad [b]=x+1,&
\eea
where $x=[X]$ is an arbitrary parameter. \par
For the fermionic generators we have
\bea
 &[\bar Q]=[Q_V]= -[Q_C]=1-z,\quad [\bar Q_V]=-[\bar Q_C]=z,
\eea 
with $z=[Q]$ an arbitrary parameter. \par
So far the parameters $\lambda, x, z$ are arbitrary. On the other hand $\lambda$ and $x$ have to be fixed by the requirement of scale and conformal invariance.  Indeed, $x$ has to be set
\bea
x&=&-\frac{1}{2}
\eea
in order to make dimensionless the free kinetic action. The parameter $\lambda$ has to be fixed
\bea\label{lambda}
\lambda&=&\frac{1}{2}
\eea
in order to guarantee the invariance under $Q_C$ of the free kinetic action. \par
Without loss of generality, the parameter $z$ can be fixed to be $z=\frac{1}{2}$ to allow $\Psi,
\bar\Psi$ having the same dimension.\par
The combinations $S\pm \bar S$ are particularly important. $S+\bar S$ is the ghost number
operator
\bea
N_{gh} &:=& S+\bar S,
\eea
while
\bea\label{scaleoperator}
D&:=& \frac{1}{2}(S-\bar S)= t\frac{d}{dt}+d_s
\eea
contains the diagonal matrix $d_s$ with the engineering or scaling dimension of the component fields. The ghost number and the scale dimensions are given by
\bea
&
\begin{array}{|c|r|r|}\hline
&N_{gh}\quad  &d_s\quad   \\\hline
X&\quad 0\quad&\quad -\demi   \quad       \\
b&\quad 0\quad& \quad     \demi  \quad \\
\Psi&\quad 1\quad & \quad  0   \quad   \\
\bar\Psi  &\quad -1\quad&\quad 0  \quad \\
\hline
\end{array}
&
\eea
The $sl(2)$ conformal algebra of the one-dimensional conformal quantum mechanics acts
with the following $D$-module unidimensional transformations on an arbitrary $s$-dimensional field $\Phi_s (t)$ (in our case $s=-\frac{1}{2}$ for $X$, $s=0$ for $\Psi$ and $\bar \Psi$, $s=\frac{1}{2}$ for $b$):
\bea
L_{-1} &=& \frac{d}{dt},\nonumber\\
L_0 ~&=& t\frac{d}{dt} + s,\nonumber\\
L_1~ &=& -t^2\frac{d}{dt} -2st.
\eea
The non-vanishing commutators are
\bea
\relax [L_0,L_{\pm 1}]&=&\pm L_{\pm1},\nonumber\\
\relax[L_1,L_{-1}]&=&2L_0.
\eea
The conformal $sl(2)$ is not a bosonic subalgebra of ${\cal G}^\sharp$. ${\cal G}^\sharp$
possesses an $sl(2)$ subalgebra given by $N_{gh}, Z, \bar Z$. This $sl(2)$ subalgebra does not
generate the conformal transformations on the component fields. On the other hand,
${\cal G}^\sharp$ possesses the  Borel subalgebra of $sl(2)$. Indeed, the subalgebra $\{ D, H\}$,
with $D$ introduced in (\ref{scaleoperator}), is identified with $\{L_0,L_{-1}\}$, so that we can identify 
$D\equiv L_0$ and $H\equiv L_{-1}$. One should note that $D$ acts on the component fields with the correct assignment of their scale dimensions.\par
It is quite remarkable, as we will discuss later, that the invariance under this subalgebra is sufficient to determine the conformally invariant actions in quantum mechanics, both for the bosonic and the fermionic sectors.\par
Even more remarkable, the invariance under just $2$ twisted fermionic generators, together with
the requirement of the vanishing of the ghost number, is sufficient to determine the superconformally invariant actions.\par
We denote as ``${\cal G}_{min}^\sharp$" the minimal ${\cal G}^\sharp$ subalgebra which,
imposed as invariance of the action, determines the full set of $N=2$ superconformal invariances. ${\cal G}_{min}^\sharp$ is given by
\bea
{\cal G}_{min}^\sharp &=&\{Q, Q_C, c, N_{gh}\}.
\eea
Imposing the ${\cal G}_{min}^\sharp$ invariance   is a very economical way to impose the full  $N=2$ superconformal invariance.\par
It is convenient to present here a list of subalgebras for ${\cal G}^\sharp$, which can be relevant
for different purposes. We have, for instance,
  \bea
  &\{Q,\bar Q_C, c\},&\nonumber\\
 & \{Q,\bar Q_C, c,  S\},&\nonumber\\
&
  \{Q,\bar Q, \bar Q_C, c,  S, H\},&\nonumber\\
& \{
  Q,\bar Q, Q_V , \bar Q_C, c,  S, H, Z\},&\nonumber\\
 &\{
  Q,\bar Q, Q_C ,\bar Q_C, c,  S, H,Z \},&\nonumber\\
 &\{
  Q,\bar Q, Q_c ,\bar Q_V,  Q_V, c,  S, H, \bar S\}.&
\eea
 As we will see in the next Section,  $Q$- and $Q_V$-invariant actions are not necessarily $Q_C$-invariant, while  $Q$- and $Q_C$-invariant actions are necessarily  $Q_V$-invariant.
  \par
An important subalgebra of ${\cal G}^\sharp$ is denoted by ${\cal B}$. Its $5$ generators are
\bea
{\cal B}\equiv
\frac{1}{2}(S-\bar S), H, Z+\bar Z, Q+Q_V-\bar Q, \bar Q-\bar Q_V-Q. \eea
As discussed in Section {\bf 6}, ${\cal B}$ coincides with the Borel subalgebra of the $sl(2|1)$ superalgebra.
\par
   After having defined  the twisted superalgebra ${\cal G}^\sharp$ with its set of nilpotent fermionic generators,  we are now looking for actions which are invariant under the full ${\cal G}^\sharp$ or some of its subalgebras. We will  restrict ourselves to the case of a standard kinetic term for the bosons $X^\mu$, namely with a Lagrangian of the form  ${\cal L}\sim \int   dt g_{\mu\nu}  \dot X^\mu   \dot X^\nu  +\ldots $.

\section{Invariant actions in the flat target-space.}
    
    \subsection {The  free supersymmetric Lagrangian.}
   
 \subsubsection {The invariance under $Q$ and $Q_V$.}
   
 Let us enforce the  $Q$ and $Q_V$  invariance of the action ${\cal S}=\int dt {\cal L}$ and let us assume the free action  to be non-dimensional.  It will therefore be uniquely defined, with the dimensionality of $X^\mu$ fixed to be $x=-\demi$.  \par
 We have 
   \bea
 {\cal S}
&=& Q Q_V\int dt \left(\bar \Psi^\mu  \eta_{\mu\nu} \Psi^\nu  \right) =  \int dt Q\left(\bar \Psi^\mu
   \eta_{\mu\nu}
   ( b^\nu-\dot X^\nu)\right)
   =\int dt \left(b^\mu
   \eta_{\mu\nu}  b^\nu
-b^\mu
   \eta_{\mu\nu}  \dot X^\nu
   +\bar \Psi^\mu  \eta_{\mu\nu}   \dot \Psi^\nu\right).\CR\nonumber\\
&&
   \eea
 By  eliminating the $b^\mu$ fields  through their algebraic equations of motion one gets  the supersymmetric Lagrangian $ {\cal L} \sim-\frac{1}{4} \dot X^\mu   \dot X^\nu    +\bar \Psi^\mu  \eta_{\mu\nu} \dot \Psi^\nu $.\par
The $Q Q_V$-exact term  $QQ_V(b^\mu  \eta_{\mu\nu} X^\nu  )$ has the appropriate dimension.  It is however,  modulo a pure time-derivative,  equal to  $QQ_V(\bar \Psi^\mu  \eta_{\mu\nu} \Psi^\nu  )$ and, therefore,  it is not independent.  Indeed,  both terms $b^\mu
   \eta_{\mu\nu}  b^\nu$ and  $-b^\mu
   \eta_{\mu\nu}  \dot X^\nu
   +\bar \Psi^\mu  \dot \Psi^\nu$ are separately $Q$-exact and thus  $Q$-invariant;  
the $Q_V$ invariance on the other hand fixes the relative coefficients of these terms. The action  
 $ \int dt ( b^\mu
   \eta_{\mu\nu}  b^\nu
-b^\mu
   \eta_{\mu\nu}  \dot X^\nu
   +\bar \Psi^\mu  \dot \Psi^\nu)$ is thus completely determined by requiring the invariance under both $Q$ and $Q_V$.

 \subsubsection {The $Q_C$ invariance.}

We leave for the time being arbitrary the parameter $\lambda$ entering (\ref{superalgebra}) and we check  
under which condition the $QQ_V$-invariant  action  $QQ_V\intdt(\bar \Psi^\mu  \eta_{\mu\nu} \Psi^\nu   )$ is also $Q_C$-invariant. We obtain 
   \bea
   Q_C{\cal L}&=&
   Q_C QQ_V\left(\bar \Psi^\mu  \eta_{\mu\nu} \Psi^\nu   \right)=
   \frac{d}{dt} ( b^\mu  \eta_{\mu\nu} X^\nu)
   +(2\lambda-1) b^\mu  \eta_{\mu\nu} \Psi^\nu.
   \eea 
   Therefore the $Q_C$-invariance is ensured  provided that $
    \lambda=\frac{1}{2} $ (see the formula (\ref{lambda})).\par
Modulo a time derivative one gets, for  the Lagrangian,
    \bea
    {\cal L}&=&Q_C \bar Q\left(  \bar \Psi^\mu  \eta_{\mu\nu}\dot  \Psi^\nu \right).
    \eea
    Therefore ${\cal L}$ is also, modulo a time derivative, $Q_C Q_V$-exact,
    \bea
    {\cal L}&=&Q_C Q_V\left( \frac{1}{2}  b^\mu  \eta_{\mu\nu}      b^\nu \right).
    \eea
    Therefore the $N=2$  free  Lagrangian is $Q,Q_V,Q_C$-invariant provided that $\lambda=\frac{1}{2}$.\par
The action  admits the following quite remarkable set of equalities
  \bea
    \int dt {\cal L} &=&  \intdt \left(b^\mu
   \eta_{\mu\nu}  b^\nu
-b^\mu
   \eta_{\mu\nu}  \dot X^\nu
   +\bar \Psi^\mu  \dot \Psi^\nu\right)\CR
   &=& \int dt Q\left(\bar \Psi ^\mu    \eta_{\mu\nu}(b^\nu-\dot X^\nu)\right)
    =\int dt QQ_V\left(\bar \Psi^\mu  \eta_{\mu\nu} \Psi^\nu  \right) = \CR&=& \int dt
      Q_C \bar Q\left(  \bar \Psi^\mu  \eta_{\mu\nu}\dot  \Psi^\nu \right)=    
    \int dt   Q_C Q_V\left( \frac{1}{2}  b^\mu  \eta_{\mu\nu}      b^\nu \right).
   \eea
  
 One can also  check the  $Q_C $ invariance as follows.
   \bea\label{qcinvcheck}
   Q_C Q_V Q\left(   \bar \Psi^\mu  \eta_{\mu\nu}  \Psi^\nu \right)
   &=& \{  Q_C, Q_V  \}Q\left(   \bar \Psi^\mu  \eta_{\mu\nu}   \Psi^\nu\right)
   +Q_V \left(Q Q_C (   \bar \Psi^\mu  \eta_{\mu\nu}\dot  \Psi^\nu )\right).\CR
   \eea
 For $\lambda=\frac{1}{2}$ the action of the bosonic symmetry  $\{Q_C,Q_V\}$  on $Q(\bar \Psi^\mu  \eta_{\mu\nu}   \Psi^\nu) $ gives zero, modulo a time-derivative.  Furthermore, for what concerns the second term in the r.h.s. of (\ref{qcinvcheck}), we have
 \bea 
 Q Q_C \left(   \bar \Psi^\mu  \eta_{\mu\nu}   \Psi^\nu \right)
 &=&Q\left( (tb^\mu -\frac{1}{2} X^\mu )\eta_{\mu\nu}   \Psi^\nu \right)=
 Q\left(  -\frac{1}{2} \Psi ^\mu \eta_{\mu\nu}   \Psi^\nu \right)=0.
   \eea
\subsection{Interactions.}
\subsubsection{The prepotential.}
The supersymmetric interaction is introduced in term of the ``prepotential"   $W[X^\mu]$.\\ 
A manifest $Q$-invariant term can be added to the action by setting
\bea
{\cal L}_{int}  & =& Q\left( \bar \Psi  ^\mu      \frac{\delta W}{   \delta X ^\mu }\right)    = b^\mu\frac{\delta W}{   \delta X ^\mu }
- \bar \Psi  ^\mu\frac{\delta^2 W}{   \delta X ^\mu \delta X ^\nu}\Psi^\nu.
  \eea
  The full $Q$-invariant action  is  thus
  \bea 
{\cal S}&=& \int dt \left(
  b^\mu
   \eta_{\mu\nu}  b^\nu
-b^\mu
   \eta_{\mu\nu} ( \dot X^\nu  +\frac{\delta W}{   \delta X ^\mu })
   +\bar \Psi^\mu  \eta_{\mu\nu}  ( \dot \Psi^\nu+
   \frac{\delta^2 W}{   \delta X ^\mu \delta X ^\nu}\Psi^\nu)\right)\nonumber
\\
      &\sim &\int dt \left(  -\frac{1}{4} \dot X^\mu   \dot X^\nu  
    -\frac{1}{4}           \frac{\delta W}{   \delta X ^\mu }      \eta_{\mu\nu}
    \frac{\delta W}{   \delta X ^\nu }
   +\bar \Psi^\mu  \eta_{\mu\nu}  ( \dot \Psi^\nu+
   \frac{\delta^2 W}{   \delta X ^\mu \delta X ^\nu}\Psi^\nu)   \right).
   \eea
   The   $Q$, $Q_V$,  $\bar Q$ and $\bar Q_V$ invariances, modulo a time derivative, of
   ${\cal L}_{int}$  are warranted because
   \bea
{\cal L}_{int}&=&
Q\bar Q \left(W\right) =Q  Q_V (W) =- \bar Q  \bar Q_V \left(W\right).
\eea    

\subsubsection{The $Q_C$ invariance.}
The $Q_C$, $\bar Q_C$  invariances, modulo a time derivative,  of  ${\cal L}_{int}$ imply the following  condition on the prepotential 
\bea
   \Psi^\mu \frac{\partial W}{   \partial X ^\nu }-
X^\nu \frac{\partial^2 W}{   \partial X ^\mu \partial X ^\nu}\Psi^\nu=0&\Rightarrow&
 \frac{\partial}{    \partial X ^\rho }\Big  (
     X^\mu \frac{\partial  W}   {   \partial X ^\mu } \Big )=0.
\eea
Therefore, the condition for having a $Q_C$-invariance  is
\bea\label{confconstr}
     X^\mu \frac{\partial  W}   {   \partial X ^\mu }&=&C,
\eea
whose general solution is
\bea
 W&=& C\ {\ln } \ R +f\left(\frac{X^ \mu}{R}\right).
\eea
$C$ is an arbitrary constant and  $f$ is an arbitrary function of the non-dimensional quantities  $\frac{X^\mu}{R}$,  where   
$R^2\equiv X^\mu \eta_{\mu\nu}  X^\nu$. \par
This gives us a so-called conformal twisted supersymmetric quantum mechanics, often  with  possible topological observables.  Its action  can be untwisted to a Lagrangian that has ordinary  conformal  supersymmetry.
For instance, for one  particle  on a plane with  coordinates $X^i, i=1,2$, and   for $f=0$,  one can   select  the conformal potential 
\bea
  \Big( \frac{\partial W}{   \partial X^i   }\Big)^2& &=   \frac{C^2}{   R ^2 }.
\eea
It defines  an interesting topological solvable   quantum mechanics on the 
${{\bf R}^2}^\ast$ plane with the origin excluded. The topological gauge function is then
$\dot X^i =  C \epsilon _{ij} \frac{   X^j}{X^2}$,
since 
\bea
\left(\dot X^i + C \epsilon _{ij} \frac{   X^j}{X^2}\right) ^2&=&
\left(\dot X^i\right)^2 + \frac{ C^2}{X^2}  +2C \ \dot\theta.
\eea
The corresponding $N=2$  supersymmetric quantum mechanics mimics as  
an elementary model the topological Yang--Mills theory, which uses the selfduality equations as  
quantum field theory topological gauge functions \cite{br}.
\par
Still keeping $i=1,2$, with $X\equiv X^1$ and  $Y\equiv X^2$,  the $Q_C$ ``topological" invariance also produces  the superconformal quantum mechanics of
a pair  of particles  evolving on a line with coordinates  $X (t)$ and $Y(t)$ and  interacting via a Calogero potential.
\\
The latter is
\bea
 \Big  (   { \frac{\partial W} {\partial X }   }  \Big) ^2  +  \Big  (   { \frac{\partial W} {\partial Y }   }  \Big) ^2 &=&
\frac{2}{(X-Y)^2}
\eea
and depends
on the relative distance.  
\par
The prepotential $W$ is indeed
\bea
W&=&\ln|X^1-X^2|= {\ln } \ R +  {\ln  }  \ |  \frac{X^1 -X^2   }{R}|.
\eea
It satisfies the general condition (\ref{confconstr}) for $Q_C$ invariance with $C=1$ 
 everywhere apart from the unphysical (infinite energy) coincidence line $X^1=X^2$  in ${\bf R}^2$.  
 
  \section{Curved target-space.}
Let us come back to the action~(\ref{cur2}), built in  \cite{bs}. 
The $\frac{1}{2}$ value of the coefficient in front of the Christoffel coefficient in the 
    ``topological gauge function" 
    $g_{\mu\nu}  
   (  \dot X^\nu) -\frac{1}{2} \Gamma _{\mu,\rho\sigma} \bar \Psi ^\rho
   \Psi^\sigma
   $
   must be fined-tuned in order to get target-space covariance.   For topological observables, defined from the cohomology of $Q$, this is not a critical issue, since they do not depend on the addition of $Q$-exact terms to the action. \par
The choice of this coefficient however, and thus the target-space covariance of the action, is implied by the additional requirement of the  $Q_V$ invariance.  Indeed, one can easily check (by using the chain rule  for the  $Q_V$ operator) that the above action  can be expressed  as a straightforward generalization of the  flat space formula,  this time with a coordinate-dependent metric $g_{\mu\nu}$. We get
 \bea
   \int dt L &= & QQ_V \intdt(\bar \Psi^\mu  g_{\mu\nu} \Psi^\nu  ).
   \eea
  Therefore,  the requirement of invariance under both  $Q$ and $Q_V$ implies the target-space covariance, which is an intriguing new result.\par   
The $Q_C$ invariance, on the other hand, is generally broken by terms which are proportional to derivatives of the metric. Thus, the compatibility between 
 conformal invariance of the worldline and the target-space covariance is broken in the presence
of a curved target-space metrics (see \cite{{bmsv},{pap}} for a discussion of the conformal invariance with non-trivial backgrounds).

 \section{Higher-order Fermi field interactions.}
 
 A series of other of $Q$ and $Q_V$ invariant actions can be constructed by simply using the  $B_{\mu_!\ldots\mu_p}(X)$ forms of various degrees. They can be written as 
  \bea
   \int dt L &= & QQ_V \intdt(\bar \Psi^\mu  \Psi^\nu  ) 
    (g_{\mu\nu}(X) +
    B_{\mu\nu} (X)
    + \sum_{p >1}
    B_{\mu\nu  \mu_1\nu_1\ldots  \mu_p\nu_p}(X)
   \bar  \Psi^{\mu_1}  \Psi^  {\nu_1} \ldots  \bar \Psi^{\mu_p}  \bar \Psi^  {\nu_p}
    ).
   \CR
  \eea
Such actions are by construction $Q$ and $Q_V$ invariant. When one expands  the above 
$Q Q_V$-exact term, one finds a $b$ dependence, such that higher Fermi field interactions do occur. Since the engineering dimensions of  $\Psi$ and $\bar\Psi$ adds to zero, there is no further limitation on the value of $p$, apart the relation $2p\leq d$, due to the fact that    $\Psi^\mu$ and $\bar\Psi^\nu$ are anticommuting fields. The forms $B_{\mu\nu}$ are analogous to the Kalb--Ramond fields.

\section{The twist transformation.}

We discuss now the relation between the twisted $N=2$ SCA ${\cal G}^\sharp$ and
the ordinary $N=2$ SCA (the $sl(2|1)$ superalgebra). Their $D$-module representations
are given in Appendix {\bf A} and {\bf B}, respectively.\par
Both superalgebras possess  a maximal common subalgebra ${\cal B}$, which  is made of 
  $5$ ($3$ even and $2$ odd) generators, defined as follows (in the right hand side of the equations we present the combinations in terms of the ${\cal G}^\sharp$ generators), 
\bea
{\cal B} &=&{\Large{\{}} D=\frac{1}{2}(S-\bar S),\quad H,\quad W=Z+\bar Z,\quad Q_1= Q+Q_V-\bar Q,\quad Q_2=\bar Q-\bar Q_V-Q{\Large{\}}}.
\eea
The superalgebra ${\cal B}$, ${\cal G}^\sharp\supset{\cal B}\subset sl(2|1)$,  is the Borel subalgebra of $sl(2|1)$ given by the Cartan and the negative-root generators.\par
One should note that no ``conformal generator" (i.e., carrying an explicit dependence on $t$)
belongs to both ${\cal G}^\sharp$ and $sl(2|1)$. This has a consequence. In order to introduce a superconformal symmetry we are faced with two
mutually exclusive paths. Either we impose the invariance under the ordinary $N=2$ superconformal algebra (ending up with an $sl(2|1)$-invariance), or we impose the invariance under the twisted generators (ending up with the $N=2$ twisted superconformal algebra ${\cal G}^\sharp$). \par
Let us denote respectively as ${\cal L}$ and ${\cal L}^\sharp$  both invariant  free Lagrangian for the ordinary $N=2$ SCA  and  the twisted $N=2$ SCA. It is convenient to introduce different notations for the component fields entering  the ordinary and the twisted $(1,2,1)$ supermultiplets.  For the
ordinary $N=2$ SCA let us have the $(Y(t);\xi_1(t),\xi_2(t);g(t))$ component fields 
and for the twisted $N=2$ SCA the $(X(t);\Psi(t),\bar\Psi(t);b(t))$ component fields.\par
With a convenient normalization we can present the free Lagrangians as
\bea
{\cal L}&=& \frac{1}{2}\left({\dot Y}^2 +g^2-\xi_1{\dot \xi_1}-\xi_2{\dot\xi_2}\right),\nonumber\\
{\cal L}^\sharp &=& b^2-b{\dot X} +\bar\Psi{\dot\Psi}.
\eea
In order to identify them (${\cal L}^\sharp={\cal L}$), we have to provide the invertible ``twist transformation" $T$,  which link both sets of fields. We have
\bea
X&=& {\sqrt 2} iY,\nonumber\\
b&=&\frac{1}{\sqrt{2}}(g+i{\dot Y}),\nonumber\\
\Psi&=&\frac{1}{{\sqrt 2}}(\xi_1-\xi_2),\nonumber\\
\bar\Psi&=&\frac{i}{\sqrt{2}}(\xi_1+\xi_2),\nonumber
\eea
or, in matrix form,
\bea
&
\begin{array}{lllllll}
T&=&\frac{1}{\sqrt{2}}\left(\begin{array}{cccc}
2i&0&0&0\\
i\partial_t&1&0&0\\
0&0&1&-1\\
0&1&i&i\end{array}\right),&\quad&
T^{-1}&=&\frac{1}{\sqrt{2}}\left(\begin{array}{cccc}
-i&0&0&0\\
-\partial_t&1&0&0\\
0&0&1&-i\\
0&0&-1&-i\end{array}\right).
\end{array}&
\eea
The above twist transformation $T$ is intrinsically complex. No solution for $T$ can be found
within the real numbers. Therefore $T$ only makes sense if the component fields $X,Y,b,g,\Psi,\bar\Psi,\xi_1,\xi_2$ are taken as complex fields. We can introduce a reality condition on the twisted fields $X, b, \Psi, \bar \Psi$ or a reality condition on the ordinary fields
$Y, g, \xi_1,\xi_2$. Both reality conditions, however, are mutually incompatible. \par
This feature of the twist transformation does not make it less useful. Indeed, 
to recover, via path integral, the ordinary correlation functions from the twisted correlation functions (or vice-versa), the only needed tool is an  analytical extension. This allows to perform the twist transformation (or its inverse). \par
In the Conclusions  we comment more on the implications of the relation between ordinary and twisted formulations for extended supersymmetric theories.\par
Let us mention here that the superalgebra ${\cal G}^\sharp$ could be further enlarged by adding
the even conformal generator $K$ (which, together with $H$ and $D$, closes the $sl(2)$ conformal algebra) and all the extra generators which are required to close the (anti)commutation relations. Such an enlarged  superalgebra   is of little significance since it does not produce any further information and constraint on the superconformal invariance, besides those obtained from
${\cal G}^\sharp$ and its relevant subalgebras. 
  
 \section{Conclusions and outlook.}
 
 In this work we proved that the $N=2$ superconformal quantum mechanics based on the $(1,2,1)$ balanced supermultiplets admits a twisted formulation, controlled by a twisted superalgebra. The twisted superalgebra ${\cal G}^\sharp$ contains $6$ nilpotent odd generators and $6$ even generators (including a central charge). The fermionic generators can be used to define BRST-type cohomologies. The invariance under different subalgebras determine different types of models. For some of them, the invariance under just a pair of the $6$ generators implies the invariance under some of the other ones.   For a curved target-space, the request of both $Q$ and $Q_V$ invariance determines, after solving the equations of motion of the auxiliary fields, a supersymmetric action which is covariant, being expressed in terms of the metric, the vectors $X$, $\Psi$, $\bar\Psi$, the covariant derivatives and the Riemann curvature.\par
A striking observation concerns the fact that the invariance  under $Q$ and $Q_C$, whose anticommutator closes on a central charge,   together with the property that one is considering Lagrangians  with  ghost number zero and standard dimension,   completely determines the superconformally actions with all their extra symmetries.\par
  These features   share many features and might be at the root of recent results in higher dimensions, for both twisted super-Yang--Mills theories~\cite {bau}  and supergravity~\cite{bbr}. 
   These quantum field cases are much more constrained, due to the presence of additional  symmetries such as the Lorentz symmetry, the R-symmetry,  etc.  The key point is that the twisted formulation  reduces in a controllable way the size of the symmetries, keeping however the symmetry algebra large  enough to uniquely determine the theory\footnote {The twisted theories possess normal local Feynmann propagators. Therefore, their simplification cannot be compared with that obtained by choosing a  light-cone gauge. In the latter case some simplicity is gained, but the local properties of the theory are harmed, due to a propagator with cut singularities which cannot be handled in a satisfactory way. Twisted theories are better behaved.}. Therefore, and quite remarkably, once the theory has been defined  by this smaller set of generators, one discovers that it possesses more symmetries, which appear ``for free". The
presence of these extra symmetries eventually allows to untwist the fermions, so that one is able to recover the physical spin-statistic relation.\par
For   the twisted $N=4$ maximal superYang-Mills theory~\cite{{vf},{mar},{bt}} one can for instance  covariantly select, among its $16$ supersymmetry generators,   $n \leq  9 $ generators \cite{top} that close ``off-shell" (twisted subalgebras can be directly constructed and regarded as a germ for  the  superPoincar\'e invariance, see \cite{{bb},{bb2}}). Moreover, only  $6$ among the $9$ generators are  needed to define the theory, either in the $N=4$, $d=4$ as  well as in the $N=1$, $d=10$ SYM cases.  Furthermore, one can enlarge the twist from the super-Poincar\'e algebra to the super-conformal case \cite{mfhs}. Within this framework one finds that   the invariance under $4$ twisted fermionic generators mixing sectors of both Poincar\'e and conformal supersymmetry   fully determines the theory~\cite{bau}. \par
In the setting of the $N=2$ supersymmetric mechanics we proved that the two  fermionic generators $Q$ and ${\bar Q}_C$, together with a central charge, contain enough information   to determine the $N=2$ superconformal action with its full set of fermionic and bosonic symmetries. We may  also mention the  recently found  example of the $N=1$, $d=4$ supergravity, which is very well-known in the untwisted formalism, and can be therefore used as a safe playground for exploring non-trivial properties of the twist procedure. It can be obtained by using only a
$U(2)\subset SO(4)$ symmetry, in analogy with the $N=1$, $d=4$ super-Yang--Mills theory.   Only $3=1+2$  global supersymmetries are needed to define,
modulo boundary terms, the supergravity.  The fourth susy generator is implied by the $3$ former ones and the full $SO(4)$ symmetry is finally found after the untwisting. \par
As a consequence of  the seemingly general existence of the twist,    burdening  quantum field theory questions,  such as the existence  of  supercharges which  only close modulo the use of some equations of motion may well  just become irrelevant since, in the twisted formulation,  one discovers that  the   theory is  determined and controlled  by a subset of supercharges which close off-shell. 
The Poincar\'e (as well as the conformal) supersymmetry might thus  appear as a kind of physical (and over-determining) effective symmetry, which sometimes emerges after untwisting a TQFT, whenever it is possible.\par
As a closing remark we wish to  point out that  we made explicit    the connection between
twisted and ordinary $N=2$ superconformal quantum mechanics, determining the twisted
superalgebra (and its $D$-module representation) associated with the twisted invariance.
A natural future step consists in investigating the relation between ordinary and twisted $N=4$ superconformal quantum mechanics. In its ``ordinary" side \cite{{dprt},{ikl},{ikl2},{kl}}, the $N=4$ superconformal algebra  (which plays the role of the $sl(2|1)$ $N=2$ SCA) is the exceptional superalgebra $D(2,1;\alpha)$. An open question is to determine its twisted superalgebra counterpart.  \par
A natural motivation to investigate twisted superconformal quantum mechanics with large $N$ comes from the geometric Langlands program which can be described \cite{kw} as a twisted $N=4$ Super-Yang--Mills theory compactified in $2D$.  It is expected that a world-line method based on the one-dimensional twisted superconformal mechanics can be employed to reconstruct the two-dimensional theory.
\par
{\quad}\par{\quad}\par
{\large{\bf Acknowledgments.}}
\par
L.B. is grateful to CBPF, where this work was completed, for hospitality and for the financial support through a PCI-BEV grant. F.T. is grateful to the LPTHE, Paris VI, for both hospitality and financial support. He acknowledges a CNPq research grant.
{\quad}\par{\quad}\par\newpage
{\Large{\bf Appendix A: the $D$-module representation of the twisted $N=2$ superconformal algebra.}}
\par
\quad\par
For completeness we present here the $D$-module representation of the twisted $N=2$ superconformal algebra ${\cal G}^\sharp$. It is a ${\bf Z}_2$-graded Lie algebra
${\cal G}^\sharp={\cal G}_0^\sharp\oplus{\cal G}_1^\sharp$, with the even sector
${\cal G}_0^\sharp$ given by the generators $H,c,S,\bar S, Z,\bar Z$ and the odd sector
${\cal G}_1^\sharp $ given by the generators $Q,\bar Q,Q_V,\bar Q_V,Q_C,\bar Q_C$.\par
The $D$-module representation consists of $4\times 4$ supermatrices acting on the supermultiplet {\small$\left(\begin{array}{c} X(t)\\ b(t)\\\Psi(t)\\\bar\Psi(t)\end{array}\right)$},
whose two upper component fields are bosonic and the two lower component fields are fermionic.
It is convenient to leave arbitrary the real parameter $\lambda$. The closure of the (anti)commutation relations is not affected by it. As discussed in the main text, see formula (\ref{lambda}), the applicability of the ${\cal G}^\sharp$ $D$-module representation to the superconformal invariance forces us to set $\lambda=\frac{1}{2}$. \par
In the expressions below we set for convenience $\partial_t:=\frac{d}{dt}$. We have
\bea
&
\begin{array}{lllllll}
Q&=&\left(\begin{array}{cccc}
0&0&1&0\\
0&0&0&0\\
0&0&0&0\\
0&1&0&0\end{array}\right),&\quad&
\bar Q&=&\left(\begin{array}{cccc}
0&0&0&1\\
0&0&0&0\\
0&-1&0&0\\
0&0&0&0\end{array}\right),
\\\nn
\end{array}&
\eea
\bea
&
\begin{array}{lllllll}
Q_V&=&\left(\begin{array}{cccc}
0&0&0&1\\
0&0&0&\partial_t\\
\partial_t&-1&0&0\\
0&0&0&0\end{array}\right),&\quad&
\bar Q_V&=&\left(\begin{array}{cccc}
0&0&1&0\\
0&0&\partial_t&0\\
0&0&0&0\\
-\partial_t&1&0&0\end{array}\right),
\\\end{array}&\nn
\eea
\bea
&
\begin{array}{lllllll}\nn
Q_C&=&\left(\begin{array}{cccc}
0&0&t&0\\
0&0&\lambda&0\\
0&0&0&0\\
-\lambda&t&0&0\end{array}\right),&\quad&
\bar Q_C&=&\left(\begin{array}{cccc}
0&0&0&t\\
0&0&0&\lambda\\
\lambda&-t&0&0\\
0&0&0&0\end{array}\right),
\\
\end{array}&
\eea
\bea
&
\begin{array}{lllllll}
H&=&\left(\begin{array}{cccc}
\partial_t&0&0&0\\
0&\partial_t&0&0\\
0&0&\partial_t&0\\
0&0&0&\partial_t\end{array}\right),&\quad&
c&=&\left(\begin{array}{cccc}
\lambda&0&0&0\\
0&\lambda&0&0\\
0&0&\lambda&0\\
0&0&0&\lambda\end{array}\right),
\\
\end{array}&
\eea
\bea
&
\begin{array}{lllllll}\nn
S&=&\left(\begin{array}{cccc}
t\partial_t-\lambda&0&0&0\\
0&t\partial_t+1-\lambda&0&0\\
0&0&t\partial_t+1-\lambda&0\\
0&0&0&t\partial_t-\lambda\end{array}\right),&\quad&
\bar S&=&\left(\begin{array}{cccc}
-t\partial_t+\lambda&0&0&0\\
0&-t\partial_t+\lambda-1&0&0\\
0&0&-t\partial_t+\lambda&0\\
0&0&0&-t\partial_t+\lambda-1\end{array}\right),
\\
\end{array}&
\eea
\bea
&\nn
\begin{array}{lllllll} 
Z&=&\left(\begin{array}{cccc}
0&0&0&0\\
0&0&0&0\\
0&0&0&1\\
0&0&0&0\end{array}\right),&\quad&
\bar Z&=&\left(\begin{array}{cccc}
0&0&0&0\\
0&0&0&0\\
0&0&0&0\\
0&0&-1&0\end{array}\right).  
\end{array}&
\eea
The (anti)-commutation relations are presented in Section {\bf 2} (formulas (\ref{anticomm}),(\ref{comm1}) and (\ref{comm2})).
\par\quad\par
%\newpage
{\Large {\bf Appendix B: the $D$-module representation of $sl(2|1)$.}}
\par
\quad\par 
In order to allow the comparison with the twisted case we present here the $D$-module representation of the $N=2$ superconformal algebra $sl(2|1)$ acting on a $(1,2,1)$ supermultiplet whose component fields have the same engineering dimensions as in the twisted case. As before, a free parameter $\lambda$ is allowed. The application of the transformations to 
superconformal invariant actions forces us to set $\lambda=\frac{1}{2}$.\par
The $D$-module representation can be obtained by closing the superalgebra recovered by applying the $sl(2)$ $D$-module representation to the $D$-module representation,
given in \cite{{pt},{krt}}, of the global $N=2$ supercharges.\par
We end up with the following set of even ($H, W, D,K$) and odd
($Q_1,Q_2,\widetilde Q_1,\widetilde Q_2$) generators
closing the (anti)commutation relations
\bea
&
\begin{array}{lll}
\relax [D,H]=-H,\quad &[D,K]=K,\quad& [K,H]=2D, \\
\end{array}
&
\eea 
together with
\bea
\{Q_1,Q_1\}=\{Q_2,Q_2\}&=&2H,\nonumber\\
\{\widetilde Q_1,\widetilde Q_1\}=\{\widetilde Q_2,\widetilde Q_2\}&=&-2K,\nonumber\\
\{Q_1,\widetilde Q_1\}=\{Q_2,\widetilde Q_2\}&=&2D,\nonumber\\
\{Q_1,\widetilde Q_2\}=\{\widetilde Q_1,Q_2\}&=&W
\eea 
 and
\bea
&
\begin{array}{ll}
\relax [H,\widetilde Q_i]=Q_i,\quad & [K,Q_i]=\widetilde Q_i,\\
\relax [ D, Q_i]= -\frac{1}{2}Q_i,\quad& [D,\widetilde Q_i]=\frac{1}{2} \widetilde Q_i,\\
\relax [ W, Q_i]= -\epsilon_{ij}Q_j,\quad& [W,\widetilde Q_i]=-\epsilon_{ij} \widetilde Q_j,\\
\end{array}&
\eea
for $i=1,2$ ($\epsilon_{12}=-\epsilon_{21}=1$).\par
Their explicit expression is given by
%\newpage
\bea
&
\begin{array}{lllllll}
H&=&\left(\begin{array}{cccc}
\partial_t&0&0&0\\
0&\partial_t&0&0\\
0&0&\partial_t&0\\
0&0&0&\partial_t\end{array}\right),&\quad&&&\\
\end{array}&\nn
\eea
\bea
&
\begin{array}{lllllll}W&=&\left(\begin{array}{cccc}
0&0&0&0\\
0&0&0&0\\
0&0&0&1\\
0&0&-1&0\end{array}\right),&&&&
\\
\end{array}&\nn
\eea
\bea
&
\begin{array}{lllllll}
D&=&\left(\begin{array}{cccc}
t\partial_t-\lambda&0&0&0\\
0&t\partial_t+1-\lambda&0&0\\
0&0&t\partial_t+\frac{1}{2}-\lambda&0\\
0&0&0&t\partial_t+\frac{1}{2}-\lambda\end{array}\right),&\quad&&&\\
\end{array}&\nn\eea
\bea
&
\begin{array}{lllllll}
K&=&\left(\begin{array}{cccc}
-t^2\partial_t+2\lambda t&0&0&0\\
0&-t^2\partial_t+(2\lambda-2)t&0&0\\
0&0&-t^2\partial_t+(2\lambda-1)t&0\\
0&0&0&-t^2\partial_t+(2\lambda-1)t\end{array}\right),&\quad&&&\\
\end{array}&\nn
\eea
\bea
&
\begin{array}{lllllll}
Q_1&=&\left(\begin{array}{cccc}
0&0&1&0\\
0&0&0&\partial_t\\
\partial_t&0&0&0\\
0&1&0&0\end{array}\right),&\quad&&&\\\nn
\end{array}&
\eea
\bea
&
\begin{array}{lllllll}\nn
Q_2&=&\left(\begin{array}{cccc}
0&0&0&1\\
0&0&-\partial_t&0\\
0&-1&0&0\\
\partial_t&0&0&0\end{array}\right),&&&&
\end{array}&\nn
\eea
\bea
&
\begin{array}{lllllll}
\widetilde Q_1&=&\left(\begin{array}{cccc}
0\quad&0\quad&t\quad&0\\
0\quad&0\quad&0\quad&t\partial_t-2\lambda+1\\
t\partial_t-2\lambda\quad&0\quad&0\quad&0\\
0\quad&t\quad&0\quad&0\end{array}\right),&\quad&&&\\
\end{array}&\nn
\eea
\bea
&
\begin{array}{lllllll}
\widetilde Q_2&=&\left(\begin{array}{cccc}
0\quad&0\quad&0\quad&t\\
0\quad&0\quad&-t\partial_t+2\lambda-1\quad&0\\
0\quad&-t\quad&0\quad&0\\
t\partial_t-2\lambda\quad&0\quad&0\quad&0\end{array}\right).
\\
\end{array}&
\eea

%\newpage

\end{document}